\documentclass{article}
\usepackage[utf8]{inputenc}

\usepackage{amsthm}
\usepackage{hyperref}
\usepackage{authblk}
\usepackage{amsmath,amstext,amssymb,amsfonts,verbatim}

\usepackage{color}

\usepackage{graphicx,algorithmic,algorithm,rotating}
\allowdisplaybreaks[1]

\newtheorem{theorem}{Theorem}
\newtheorem{definition}[theorem]{Definition}

\newenvironment{myenumerate}{
\begin{enumerate}
 \setlength{\itemsep}{1pt}
 \setlength{\parskip}{0pt}
 \setlength{\parsep}{0pt}}{\end{enumerate}
}

\newenvironment{myitemize}{
\begin{itemize}
 \setlength{\itemsep}{1pt}
 \setlength{\parskip}{0pt}
 \setlength{\parsep}{0pt}}{\end{itemize}
}

\title{\vspace{-4cm}The Thermodynamics of Network Coding, and an Algorithmic Refinement of the Principle of Maximum Entropy}

\begin{document}

\author[1,2,3,4,5]{Hector Zenil\thanks{An animated video explaining some of the methods is available at: \url{https://www.youtube.com/watch?v=BEaXyDS_1Cw} Corresponding author: hector.zenil [at] algorithmicnaturelab [dot] org}}
\author[1,2,3]{Narsis A. Kiani}
\author[2,5]{Jesper Tegn\'er}
\affil[1]{Algorithmic Dynamics Lab$^2$}
\affil[2]{Unit of Computational Medicine, Center for Molecular Medicine, Department of Medicine Solna, Karolinska Institute, Stockholm, Sweden.}
\affil[3]{Algorithmic Nature Group, LABORES for the Natural and Digital Sciences, Paris, France}
\affil[4]{Oxford Immune Algorithmics, Oxford, U.K.}
\affil[5]{Biological and Environmental Sciences and Engineering Division,\\Computer, Electrical and Mathematical Sciences and Engineering\\Division, King Abdullah University of Science and\\ Technology (KAUST), Kingdom of Saudi Arabia.}
\date{}

\maketitle

\vspace{-1cm}
\begin{abstract}
The principle of maximum entropy (Maxent) is often used to obtain prior probability distributions as a method to obtain a Gibbs measure under some restriction giving the probability that a system will be in a certain state compared to the rest of the elements in the distribution. Because classical entropy-based Maxent collapses cases confounding all distinct degrees of randomness and pseudo-randomness, here we take into consideration the generative mechanism of the systems considered in the ensemble to separate objects that may comply with the principle under some restriction and whose entropy is maximal but may be generated recursively from those that are actually algorithmically random offering a refinement to classical Maxent. We take advantage of a causal algorithmic calculus to derive a thermodynamic-like result based on how difficult it is to reprogram a computer code. Using the distinction between computable and algorithmic randomness we quantify the cost in information loss associated with reprogramming. To illustrate this we apply the algorithmic refinement to Maxent on graphs and introduce a Maximal Algorithmic Randomness Preferential Attachment (MARPA) Algorithm, a generalisation over previous approaches. We discuss practical implications of  evaluation of network randomness. Our analysis provides insight in that the reprogrammability asymmetry appears to originate from a non-monotonic relationship to algorithmic probability. Our analysis motivates further analysis of the origin and consequences of the aforementioned asymmetries, reprogrammability, and computation.\\ 

\end{abstract}

\begin{quote}
    Thermodynamics is a funny subject. The first time you go through it, you don't understand it at all. The second time you go through it, you think you understand it, except for one or two small points. The third time you go through it, you know you don't understand it, but by that time you are so used to it, it doesn't bother you any more.
    
\flushright  Arnold Sommerfeld~\cite{angrist}.
\end{quote}

\section{Classical Thermodynamics and Related Work}

We propose a conceptual framework and a set of heuristics and methods for object randomization based on computability and algorithmic probability different to those from classical methods that offer a refinement on the \textit{Principle of Maximum Entropy} and also a different perspective of thermodynamics of computation from the point of view of algorithmic information.

Conventionally, and traditionally, thermodynamic entropy is defined as follows:

\begin{myitemize}
\item A measure of \textit{statistical} disorder;
\item Some quantity or property that increases but never decreases;
\item A process that defines the direction of time;
\item A measure of \textit{statistical} information
\end{myitemize}

Some of the problems surrounding the characterisation of the second law and entropy go back about a hundred years, to the time when they were introduced. While most of the discussion around thermodynamics is not only legitimate but central to the most pressing and important questions in physics, the statistical version of entropy has found a renewed application in the form of the so-called \textit{Principle of Maximum Entropy}, often denoted by \textit{Maxent}.

Previous work has considered the question of replacing all or part of the statistical machinery from statistical mechanics in order to arrive at an algorithmic approach to thermodynamics. Some authors have discussed the analogy between
algorithms and entropy~\cite{grunwald,hammer}. One example is the thought experiment  `Maxwell's demon'--whereby
Maxwell suggested that the second
law of thermodynamics might hypothetically be violated by introducing intelligence in the form of a being capable of following an algorithm that enabled it to distinguish between particles of high and low energy--taken together with Szilard's discussion of this paradox~\cite{szilard}. Here we offer an alternative framework 
in terms of which Maxwell's paradox and other thermodynamic-like phenomena can be
reformulated in terms of computer programs via algorithmic probability and program-size complexity.

Other examples combining computation and entropy can be found in Seth Lloyd's concept of \textit{thermodynamic depth}~\cite{lloyd}, heavily indebted to the work of Kolmogorov and Chaitin and to Bennett's notion of \textit{logical depth}~\cite{bennett}; in Baez's \textit{algorithmic thermodynamics}\cite{baez} approach that is capable of defining an algorithmic version of \textit{algorithmic temperature} and \textit{algorithmic
pressure}, and in Crutchfield's \textit{computational mechanics} using \textit{epsilon-machines}~\cite{crutchfield}. Zurek has also proposed the inclusion of algorithmic randomness in a definition of physical entropy to allow the reformulation of aspects of thermodynamics, such as the Maxwell demon's thought experiment~\cite{zurek1,zurek2}.

The interest in introducing algorithmic complexity to questions of thermodynamics and the second law derives from a wish to introduce an additional dimension into the discussion of the foundations of probability going back to Kolmogorov~\cite{kolmo}, Solomonoff~\cite{solomonoff}, Chaitin~\cite{chaitin} and Levin~\cite{levin} among others, but also 
from an interest in the relationship between physics and logical information~\cite{baez,bennett,grunwald,zurek1,zurek2,szilard,lloyd}. This include, testing or refining the second law of thermodynamics by taking into consideration and ruling out apparent disordered states that are not algorithmically random. Unlike statistical mechanical approaches, algorithmic complexity represents a generalisation over entropy that assigns lower entropy to objects that not only appear statistically simple but are also algorithmically simple by virtue of having a short generating mechanism capable of reproducing the algorithmic content of a system. Without such an additional dimension, uncomputable or algorithmically random and computable and non-algorithmically random networks are conflated and collapsed into the typical Bernoulli distribution produced by entropy, in which maximal entropy represents apparent statistical disorder, without distinguishing between networks with an algorithmic construction and algorithmic randomness. Indeed, a random-looking system with maximal entropy can be recursively generated by a short computer program that entropy would classify as statistically random but not as algorithmically random. An example of a fundamental limitation of Shannon entropy is, for example, offered in~\cite{zkpaper}, where its fragility is exposed in a very simple example involving an attempt to quantify the deterministic vs random nature of exactly the same object (a recursive graph of algorithmic randomness).

While we will demonstrate that the mathematics of changes in algorithmic complexity and the formal repurposing capabilities of computer programs show a \mbox{thermodynamic-like} phenomenon that is similar, if not equivalent, to the mathematics of physical thermodynamics, in this paper we do not aim to connect physical thermodynamics to algorithmic thermodynamics directly. In other words, while we may be able to count the number of operations for converting one computer program into another by targeting a specific desired output from these programs, we do not enter in this paper into the details of the energetic cost of implementing these operations in hardware. We believe, however, that these results, while abstract in nature, are fundamentally `analogous' in character
to the same kind of thermodynamic principles in classical mechanics. While this is beyond the scope of the present paper, we are motivated by the current results to revisit this important topic.

\section{Notation and Definitions}

Let $G$ be a graph composed of a set $E(G)$ of edges with end points in $V(G)$ also known as the nodes of $G$. We will denote by $|V(G)|$ the cardinality of $V(G)$ and by $|E(G)|$ the cardinality of the set $E(G)$.

\begin{definition}
The degree of a node in $G$ is the number of directed and undirected edges linked to that node. 
\end{definition}

\begin{definition}
The degree sequence of a graph is the unsorted list of degree nodes of all nodes of $G$.
\end{definition}
Generally speaking random graph is described  by a probability distribution, or by a random process which generates them. Very intuitively, a random graph is obtained by starting with a set of number of  isolated vertices and adding successive edges between them at random. Diverse models has been suggested to generate A random graph \cite{erdos,erdos2, gilbert}. Each of these model produce different probability distributions on graphs. Most commonly studied model,  Erd\"os-R\'enyi model assigns equal probability to all graphs with equal number of edges, i.e. In  Erd\"os-R\'enyi or E-R graph, each edge is included in the graph with probability $p$ independent from every other edge.

The parameter $0\geq p \leq 1$ determines the graph density and as it increases the model becomes more and more likely to include graphs with more edges. The value $p = 0.5$ corresponds to the case where all graphs on $n$ vertices are chosen with equal probability and degree sequence approximating a uniform distribution, i.e. almost surely most nodes have about the same node degree.
An E-R random graph can be obtained by starting with a set of $n$ isolated nodes and adding random edges between them. Let $\phi$ be a property of $G$. We can quantify the  probability of graph $G$ to have that particular property $Prob(\phi(G))$ based on the set of elements from the distribution of the mathematical model used to produce the random graph. Generally the  probability of property $\phi$ can take values between 0 and 1 as $V(G)$ or $E(G)$ tends to infinity. These models can be used in the probabilistic method to provide a  definition of what it means for a property to hold for almost all graphs. We can speak of the set of first-order properties of a graph for which the previous value is 1, rather than 0; let us refer to such properties as ``holding almost surely". In the case of properties of E-R graphs, we will always mean properties that ``hold almost surely" and which can be satisfied by graphs on at most $|V(G)|$ vertices. We will often shorten it to read '$G$ has this property' meaning ``sharing many of the well-known properties that hold almost surely for ER graphs" and we will refer to  Erd\"os-R\'enyi-like graphs as E-R graphs.

\begin{definition}
 An object $\alpha$ is algorithmic random if there exists a constant c such that for all $n$,
 $K(\alpha | n)\leq n-c $, where $\alpha | n $ denotes the first $n$ bits of $\alpha$. 
 i.e.  an algorithmic random object is almost algorithmically incompressible. \cite{levin,chaitin}
\end{definition}

\begin{definition}
We define the elements of an object $G$ that are able to `move' $G$ towards algorithmic randomness (i.e. find the elements of the original graph that can produce another more or less algorithmic random graph) as $N(G)$ standing for the set of elements that can transform $G'$ into a more random graph by at least $log_2(|G|)$ bits; and as $P(G)$ the set of elements that make $G$ away from randomness.
\end{definition}

\begin{definition}
The complementary set of $N(G) \cup P(G)$ is the set of neutral elements and are those that upon removal they do not change $G$ neither towards greater or lower algorithmic randomness. If not otherwise specified, the neutral set will also contain the elements that `move' $G$ towards or away from algorithmic randomness by less than a constant $log_2(n) + c$, with $c$ a small positive real number~\cite{nature,algodyn}.
\end{definition}

\begin{definition}
We call the ordered set of all elements of $G$, $\sigma_G$, and their algorithmic contribution values sorted from most positive to most negative contribution to the algorithmic content of $G$ and to $\sigma_N(G)$ and $\sigma_P(G)$, the negative and positive parts, i.e. the ordered sets of $N(G)$ and $P(G)$.
\end{definition}

\noindent We call $\sigma_(G)$ the `signature' of $G$.

\begin{definition}
We call a deletion an \textit{algorithmically random deletion} (or simply a deletion if not otherwise stated), if it is non-recursive, non-computable or non-enumerable (which will be used as synonyms ad opposing computable, algorithmic or recursive), that is, it cannot be algorithmically recovered or if its information was not recorded by, e.g., an index from an enumeration of elements of $G$ when deleted, and is thus, for all computable purposes, lost.
\end{definition}

\begin{definition}
An algorithmically random graph is a graph whose $P(G)$ is empty, that is, it has no elements that can move it away from randomness. 
\end{definition}

\begin{definition}
An algorithmically random Erd\"os-R\'enyi (E-R) graph is an E- R graph with $|P(G)|=0$
\end{definition}

\noindent Not all E-R graphs are algorithmically random (see Theo.~\ref{theoexist}).\\

\subsection{Graph Entropy}
\label{entropy}

A major challenge in science is to provide proper and suitable representations of graphs and networks for use in fields ranging from social sciences to physics~\cite{boccaletti} and chemistry~\cite{chen2014entropy}.

Networks have been characterized using classical information theory. One problem in this area is the interdependence of many graph-theoretic properties, which makes measures more sophisticated than single-property measurements~\cite{orsini} difficult to come by. The standard way to address this is to generate graphs that have a certain specific property while being random in all other respects, in order to check whether or not the property in question is typical among an ensemble of graphs with otherwise seemingly different properties. 
The Shannon entropy (or simply entropy) of a graph $G$ can be defined by 

$$H_i(G)=-\sum_{i=1}^n P(G(x_i)) \log_2 P(G(x_i)) \textnormal{ (Eq. 2)}$$

\noindent where $G$ is the random variable with property $i$ and $n$ possible outcomes (all possible adjacency matrices of size $|V(G)|$). For example, a completely disconnected graph $G$ with all adjacency matrix entries equal to zero has entropy $H(G)=0$, because the number of different symbols in the adjacency matrix is 1. However, if a different number of 1s and 0s occur in $G$, then $H(G)\neq0$.

Just as for strings, Shannon entropy can also be applied to the node degree sequence of a graph~\cite{korner1988random}.
The Shannon entropy of an unlabelled network characterized by its degree distribution can be described by the same formula for Shannon entropy where the random variable is a degree distribution. The chief advantage of so doing is that it is invariant to relabellings. This also means that the degree distribution is not a lossless representation of a labelled network (but rather of its isomorphic group), and is an interesting entropic measure, but one that can only be used when the node labels are not relevant. For example, in a causal recursive network, the node labels may represent time events in a sequence that has a meaning captured in the network labelling, in which case the degree distribution sequence (where no agreement has been reached on the order of the elements in the distribution sequence, which is therefore disordered) cannot be used to reconstruct the original network represented by the unlabelled version or the isomorphism group of the labelled networks.
It is also clear that the concept of entropy rate cannot be applied to the degree distribution, because the node degree sequence has no particular order, or any order is meaningless because any label numbering will be arbitrary. This also means that Shannon entropy is not invariant to the language description of a network, especially as a labelled or an unlabelled network, except for clearly extreme cases (e.g. fully disconnected and completely connected networks, both of which have flat degree distributions and therefore the lowest Shannon entropy for degree sequence and adjacency matrix).
While the application of Entropy to graph degree distributions has been relatively more common, the same Entropy has also been applied to other graph features, such as functions of their adjacency matrices~\cite{estrada2014walk}, and to distance and Laplacian matrices~\cite{Dehmer3}. A survey contrasting adjacency matrix based (walk) entropies and other entropies (e.g. on degree sequence) is offered in~\cite{estrada2014walk}. It finds that adjacency based ones are more robust vis-\'a-vis graph size and are correlated to graph algebraic properties, as these are also based on the adjacency matrix (e.g. graph spectrum). 
In general we will use Block entropy in order to detect more graph regularities (through the adjacency matrix) at a greater resolution. But for Block entropy there is an extra factor to be taken into account. The unlabelled calculation of the Block entropy (not relevant for 1-bit entropy) of a graph has to take into consideration all possible adjacency matrix representations for all possible labellings. Therefore, the Block entropy of a graph is given by:

$$
H(G)=\min\{H(A(G_L)) | G_L \in L(G)\}
\textnormal{ (Eq. 3)}$$

\noindent where $L(G)$ is the group of all possible labellings of $G$. 

In~\cite{zkpaper}, we introduced a graph that is generated recursively by a small computer program of (small) fixed length, yet when looking at its degree sequence it tends to maximal entropy and when looking at the adjacency matrix it tends to zero entropy at the limit, thus displaying divergent values for the same object when considering different mass probability distributions---when
assuming the uniform distribution to characterize an underlying ensemble comprising all possible adjacency matrices of increasing size, or when assuming all possible degree sequences in the face of a total lack of knowledge of the deterministic nature of the graph in question.

\section{Algorithmic Information Dynamics}

The expected value of algorithmic entropy equals its Shannon entropy up to a constant that depends only on the distribution~\cite{antunes}. That is, every deterministic source has both low entropy and low algorithmic randomness, and algorithmically random objects surely have the highest Shannon entropy. However, in practical terms, they are fundamentally different. Nowhere in Shannon entropy is there any indication as to how to estimate the underlying mass probability distribution needed to determine the random or deterministic nature of a source, the availability of some other method for doing so being simply assumed. 

Algorithmic complexity, however, does provide many methods, albeit very difficult ones, to estimate the algorithmic randomness of an object by inspecting the set of possible programs whose size is at most the size of the shortest program that may produce the object. One popular way to approximate it has been by using lossless compression algorithms, given that a compressed program is sufficient proof of non-randomness.

The algorithmic (Kolmogorov-Chaitin) complexity of a string $x$ is the length of the shortest effective description of $x$. There are several versions of this notion. Here we use mainly the plain complexity, denoted by $C(x)$. 

We work over the binary alphabet $\{0, 1\}$. A string is an element of $\{0, 1\}^*$. If $x$ is a string, $|x|$ denotes its length. Let $M$ be a universal Turing machine that takes two input strings and outputs one string. For any strings $x$ and $y$, we define the algorithmic complexity of $x$ conditioned by $y$ with respect to $M$, as:

$$C_M (x | y) = \min\{|p| \textit{ such that } M(p, y) = x\}.$$

We match the machine $M$ with a universal machine $U$, thereby allowing us to drop the subscript. We then write $C(x | y)$ instead of $C_M (x | y)$. We will also write $C(x)$ instead of $C(x | \lambda)$ (where $\lambda$ is the empty string).

\subsection{Approximations to Algorithmic Complexity}

we have introduced methods to approximate the algorithmic complexity of a graph with interesting results~\cite{zenilgraph,zenilkiani,journalcomplexnetworks}. For example, in \cite{zenilgraph} correlations were reported among algebraic and topological properties of synthetic and biological networks by means of algorithmic complexity, and an application to classify networks by type was developed in \cite{zenilkiani}. Together with~\cite{zenilgraph} and~\cite{zenilkiani}, the methods introduced represented a novel view and constitute a formal approach to graph complexity, while providing a new set of tools for the analysis of the local and global structures of networks.

The algorithmic probability~\cite{solomonoff,levin,kirchherr} of a string $s$, denoted by $AP(s)$ provides the probability that a valid random program $p$ written in bits uniformly distributed produces the string $s$ when run on a universal (prefix-free~\footnote{The group of valid programs forms a prefix-free set (no element is a prefix of any other, a property necessary to keep $0 < AP(s) < 1$.) For details see~\cite{cover,calude}.}) Turing machine $U$. Formally,

$$AP(s) = \sum_{p:U(p) = s} 1/2^{|p|} \textnormal{ (Eq. 5)}$$

That is, the sum over all the programs $p$ for which a universal Turing machine $U$ outputs $s$ and halts.

Algorithmic probability and algorithmic complexity $K$ are formally (inversely) related by the so-called algorithmic Coding theorem~\cite{cover,calude}:

$$|-\log_2 AP(s) - K(s)| < \mathcal{O}(1) \textnormal{ (Eq. 6)}$$

As shown in~\cite{zenilgraph}, estimations of algorithmic complexity are able to distinguish complex from random networks (of the same size, or growing asymptotically), which are both in turn distinguished from regular graphs (also of the same size). $K$ calculated by the BDM assigns low algorithmic complexity to regular graphs, medium complexity to complex networks following Watts-Strogatz or Barab\'asi-Albert algorithms, and higher algorithmic complexity to random networks. That random graphs are the most algorithmically complex is clear from a theoretical point of view: nearly all long binary strings are algorithmically random, and so nearly all random unlabelled graphs are algorithmically random~\cite{randomgraphs}, where algorithmic complexity is used to give a proof of the number of unlabelled graphs as a function of its randomness deficiency (how far it is from the maximum value of $K(G)$).

The \textit{Coding Theorem Method} (CTM)~\cite{d4,d5} is rooted in the relation specified by algorithmic probability between frequency of production of a string from a random program and its algorithmic complexity (Eq.~(6). It is also called the algorithmic \textit{Coding theorem}, to contrast it with another coding theorem in classical information theory). Essentially it uses the fact that the more frequent a string (or object), the lower its algorithmic complexity; and strings of lower frequency have higher algorithmic complexity.

\subsection{Block Decomposition Method}

The \textit{Block Decomposition Method} (BDM) was introduced in~\cite{zenilgraph} and ~\cite{bdmpaper}. It requires the partition of the adjacency matrix of a graph into smaller matrices using which we can numerically calculate its algorithmic probability by running a large set of small 2-dimensional deterministic Turing machines, and then-- by applying the algorithmic Coding theorem-- its algorithmic complexity. Then the overall complexity of the original adjacency matrix is the sum of the complexity of its parts, albeit with a logarithmic penalization for repetition, given that $n$ repetitions of the same object only add $\log n$ to its overall complexity, as one can simply describe a repetition in terms of the multiplicity of the first occurrence. More formally, the algorithmic complexity of a labelled graph $G$ by means of $BDM$ is defined as follows:

\begin{equation}
\label{newecaeq}
BDM(G,d) = \sum_{(r_u,n_u)\in A(G)_{d\times d}} \log_2(n_u)+K_m(r_u)
\end{equation}
where $K_m(r_u)$ is the approximation of the algorithmic complexity of the subarrays $r_u$ arrived at by using the algorithmic Coding theorem (Eq.~(6)), while $A(G)_{d\times d}$ represents the set with elements $(r_u,n_u)$, obtained by decomposing the adjacency matrix of $G$ into non-overlapping squares of size $d$ by $d$. In each $(r_u,n_u)$ pair, $r_u$ is one such square and $n_u$ its multiplicity (number of occurrences). From now on $BDM(g,d=4)$ will be denoted only by $K(G)$, but it should be taken as an approximation to $K(G)$ unless otherwise stated (e.g. when taking the theoretical true $K(G)$ value). Once CTM is calculated, BDM can be implemented as a look-up table, and hence runs efficiently in linear time for non-overlapping fixed size submatrices.

\subsection{Normalized BDM}
\label{norm-bdm-sec}

A normalized version of BDM is useful for applications in which a maximal value of complexity is known or desired for comparison purposes. The chief advantage of a normalized measure is that it enables a comparison among objects of different sizes, without allowing size to dominate the measure. This will be useful in comparing arrays and objects of different sizes. First, for a square array of size $n\times n$, we define:

\begin{equation}
Min BDM(n)_{d\times d} = \lfloor n/d \rfloor + \displaystyle\min_{x\in M_d(\{0,1\})} CTM(x)
\end{equation}

\noindent where $M_d(\{0,1\})$ is the set of binary matrices of size $d\times d$. For any $n$, $MinBDM(n)_{d\times d}$ returns the minimum value of Eq.~\eqref{newecaeq} for square matrices of size $n$, so it is the minimum BDM value for matrices with $n$ nodes. It corresponds to an adjacency matrix composed of repetitions of the least complex $d\times d$ square. It is the all-1 or all-0 entries matrix, because $0_{d,d}$ and $1_{d,d}$ are the least complex square base matrices (hence the most compressible) of size $d$.

Secondly, for the maximum complexity, Eq.~\eqref{newecaeq} returns the highest value when the result of dividing the adjacency matrix into the $d\times d$ base matrices contains the highest possible number of different matrices (to increase the sum of the right terms in Eq.~\eqref{newecaeq}) and the repetitions (if necessary) are homogeneously distributed along those squares (to increase the sum of the left terms in Eq.~\eqref{newecaeq}) which should be the most complex ones in $M_d(\{0,1\})$. For $n,d\in \mathbb{N}$, we define a function \[f_{n,d}: M_d(\{0,1\})\longmapsto \mathbb{N}\]
that verifies:
\begin{equation}
  \label{eq:2}
   \displaystyle\sum_{r\in M_d(\{0,1\})} f_{n,d}(r) = \lfloor
  n/d\rfloor^2  \\
\end{equation}
  \begin{equation}
  \label{eq:3}
   \displaystyle\max_{r\in M_d(\{0,1\})} f_{n,d}(r)\\ \leq\ \ 1+
  \displaystyle\min_{r\in M_d(\{0,1\})} f_{n,d}(r)   \\
 \end{equation}
 \begin{equation}  
 \label{eq:4}
  CTM(r_i) > CTM(r_j)\ \Rightarrow \ f_{n,d}(r_i)\geq
  f_{n,d}(r_j)
\end{equation}

The value $f_{n,d}(r)$ indicates the number of occurrences of $r\in
M_d(\{0,1\})$ in the decomposition into $d\times d$ squares of the
most complex square array of size $n\times n$. Condition Eq.~\eqref{eq:2}
establishes that the total number of component squares is $\lfloor
n/d\rfloor^2$. Condition Eq.~\eqref{eq:3} reduces the square repetitions 
as much as possible, so as to increase the number of differently composed squares as far as possible and distribute them homogeneously. Finally, Eq.~\eqref{eq:4} ensures that the most complex squares are the best represented. Then, we define:
\[
Max BDM(n)_{d\times d} = \hspace{-0.5cm}\sum_{
  {\begin{array}{c}
    r\in M_d(\{0,1\}),\\f_{n,d}(r)>0
  \end{array}}} \hspace{-0.5cm}
\log_2(f_{n,d}(r))+ CTM(r)
\]
Finally, the normalized BDM value of an array $X$ is:\\

Given a square matrix $X$ of size $n$,
$NBDM(X)_d$ is defined as 
\begin{equation}
\label{nbdm}
\frac{CTM(X) - Min BDM(n)_{d\times
    d}}{Max BDM(n)_{d\times d} - Min BDM(n)_{d\times d}}
\end{equation}
This way we take the complexity of an array $X$ to have a normalized value which is not dependent on the size of $X$ but rather on the relative complexity of $X$ with respect to other arrays of the same size. The use of $Min BDM(n)_{d\times d}$ in the normalization is relevant. Note that the growth of $Min BDM(n)_{d\times d}$ is linear with $n$, and the growth of $Max BDM(n)_{d\times d}$ exponential. This means that for high complexity matrices, the result of normalizing by using just $CTM(X)/Max BDM(n)_{d\times d}$ would be similar to $NBDM(X)_d$. But it would not work for low complexity arrays, as when the complexity of $X$ is close to the minimum, the value of $CTM(X)/Max BDM(n)_{d\times d}$ drops exponentially with $n$. For example, the normalized complexity of an empty array (all 0s) would drop exponentially in size. To avoid this, Eq.~\eqref{nbdm} considers not only the maximum but also the minimum. 

Notice the heuristic character of $f_{n,d}$. It is designed to ensure a quick computation of $Max BDM(n)_{d\times d}$, and the distribution of complexities of squares of size $d\in\{3,4\}$ in $D(5,2)$ ensures that $Max BDM(n)_{d\times d}$ is actually the maximum complexity of a square matrix of size $n$, but for other distributions it could work in a different way. For example, condition~\eqref{eq:3} assumes that the complexities of the elements in $M_d(\{0,1\})$ are similar. This is the case for $d\in\{3,4\}$ in $D(5,2)$, but it may not be true for other distributions. But at any rate it offers a way of comparing the complexities of different arrays independent of their size.

\subsection{A Reprogrammability Calculus}

At the core of the (re-)programmability analysis is a causal calculus introduced in~\cite{maininfo} based on Algorithmic Information Dynamics, the change in complexity of a system subject to perturbations, particularly the direction (sign) and magnitude of the change in algorithmic information content $C$  
between wild and perturbed instants of the same object, such as objects $G$ and $G'$, which for purposes of illustration can be graphs with a set of nodes $V(G)$ and a set of edges $E(G)$.

We define two measures of reprogrammability (denoted by $P_R(G)$ and $P_A(G)$ as introduced in~\cite{maininfo}) as measures that capture different aspects of the capability of a system's elements to move an object $G$ towards or away from randomness.

\begin{definition}
Relative programmability is defined as $P_R(G) := MAD(\sigma(G))) / n$ or 0 if $n = 0$, where $n = \max{\{|\sigma(G)|\}}$ measures the shape of $\sigma_P(G)$ and how it deviates from other distributions (e.g. uniform or normal), and MAD is the median absolute deviation of $G$.
\end{definition}

\begin{definition}
Absolute programmability is defined as $P_A(G) := |S(\sigma_P(G)) - S(\sigma_N(G)) | / m$, where $m = \max(S(\sigma_P(G)), S(\sigma_N(G)))$ and $S$ is an interpolation function. This measure of programmability captures not only the shape of $\sigma_P(G)$ but also the sign of $\sigma_P(G)$ above and below $x = 0$.
\end{definition}

The dynamics of a graph can then be defined as transitions between different states, and one can always ask after the potential causal relationship between $G$ and $G'$. In other words, what possible underlying minimal computer program can explain $G'$ as evolving over discrete time from state $G$?
 
For graphs, we can allow the operation of edge $e$ removal from $G$ denoted by $G\backslash e$ where the difference $| C(G) - C(G\backslash e) |$ is an estimation of the (non-)shared algorithmic mutual information of $G$ and $G\backslash e$ or the \textit{algorithmic information dynamics} (or \textit{algorithmic dynamics} in short) for evolving time-dependent systems (e.g. if $G'$ evolves from $G$ after $t$ steps). If $e$ does not contribute to the description of $G$, then $| C(G) - C(G\backslash e) | \sim \log_2|V(G)|$, where $|V(G)|$ is the node count of $G$, i.e. the algorithmic dynamic difference will be very small and at most a function of the graph size, and thus the relationship between $G$ and $G'$ can be said to be causal and not random, as $G'$ can be derived from $G'$ with at most $\log_2|V(G)|$ bits. If, however, $| C(G) - C(G\backslash e) | > \log_2|V(G)|$ bits, then $G$ and $G\backslash e$ are not states having the same causal origin and the algorithmically random removal of $e$ results in a loss because $G\backslash e$ cannot recursively/computably recover $G$ and has to be explained independently, e.g. as noise. In contrast, if $C(G) - C(G\backslash e) < \log_2|V(G)|$, then $e$ can be explained by $G$ alone and it is algorithmically not contained/derived from $G$, and is therefore in the causal trajectory of the description of $G$ from $G\backslash e$ and of $G\backslash e$ from $G$. 

If $G$ is random, then the effect of $e$ will be small in either case, but if $G$ is richly causal and has a very small generating program, then $e$ as noise will have a greater impact on $G$ than would removing $e$ from the description of an already short description of $G$. However, if $| C(G) - C(G\backslash e) | \leq \log_2 |V(G)|$, where $|V(G)|$ is the vertex count of $G$, then $e$ is contained in the algorithmic description of $G$ and can be recovered from $G$ itself (e.g. by running the program from a previous step until it produces $G$ with $e$ from $G\backslash e$).

\section{The Thermodynamics of Computer Programs}

A thermodynamic-like result is illustrated by a measure of \textit{sophistication} based on quantifying the difficulty of reprogramming an object according to its underlying algorithmic generator or
computer program. A measure of sophistication is a measure capable of telling apart `sophisticated' cases from simple and random objects, the latter being assigned low complexity, as in the case of Bennett's logical depth~\cite{bennett}.

For example, in a complete graph $G$, the algorithmic-random deletion of any single node or single edge has the same effect, because all nodes and all edges make the same algorithmic-information contribution to the original graph as they cannot be distinguished in any way, and so their recovery does not need recording in, e.g., an enumerable index. In this case, for a complete graph, the ordered elements $\sigma(G)$ can be analytically derived from the uniform mass probability distribution defined by $x = \log_2 |V(G)|$ with $|V(G)|$ the node count of $G$ because all nodes in a complete graph are neutral since they contribute by $\log_2 |V(G)|$. In contrast, the algorithmically random deletion of any single edge does not lead to a complete graph, and after 2 edge deletions, recovering any single one is not algorithmic, as it entails a lack of recording. All edges are therefore in $N(G)$ as they move the complete graph $G$ towards algorithmic randomness.

\begin{figure}[ht]
\centering
\includegraphics[width=10cm]{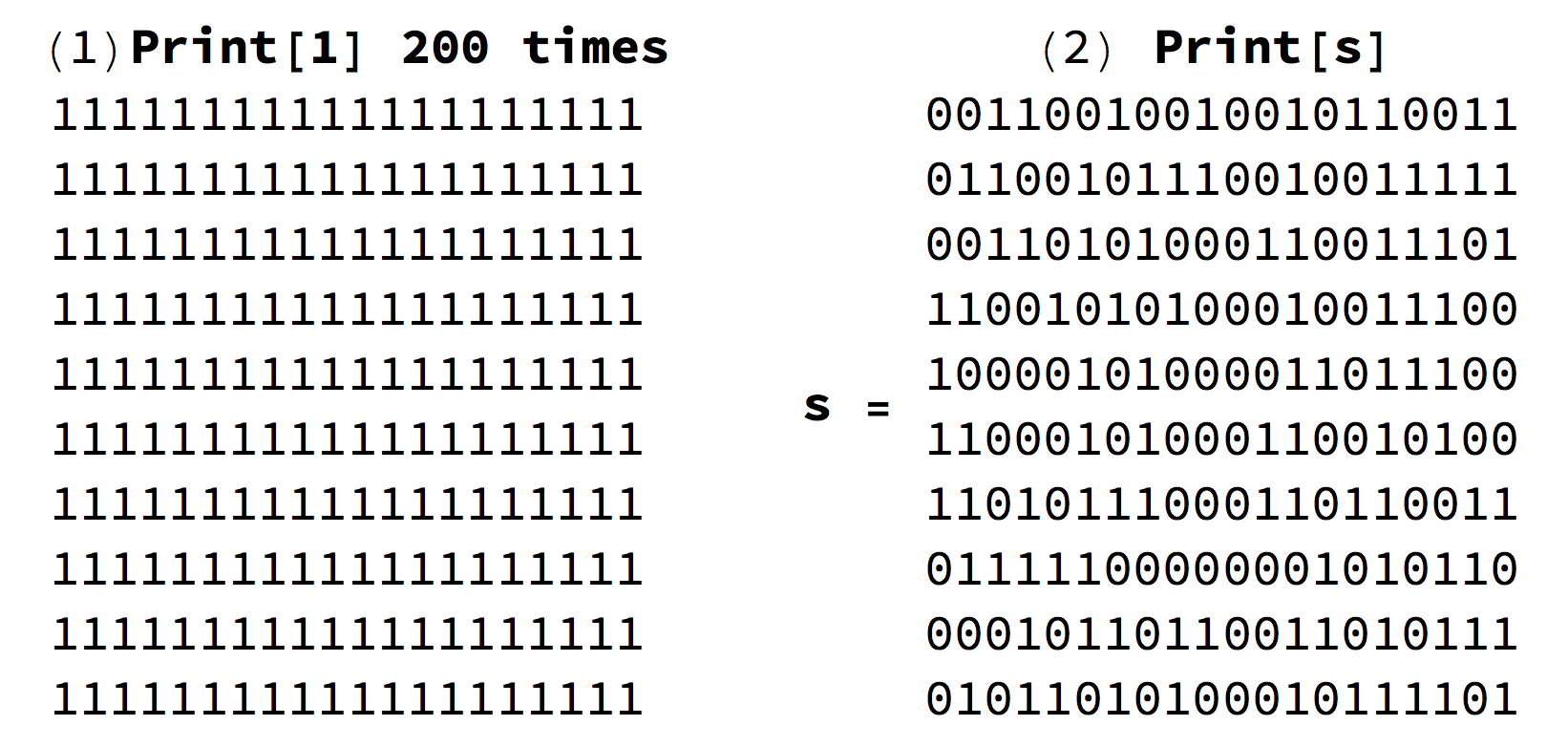}

\caption{The thermodynamics of computer programming. Top: The programs producing simple versus random data have different reprogrammability properties. If repurposed to generate programs to print blocks of 0s, we only need a single intervention in the generative program of (1), changing 1 to 0 inside the \textbf{Print} instruction indicating that 200 0s be printed instead of 1s. In contrast, asking a program that prints a random binary string $s$ to print only 0s will require on average $|s|/2$ interventions to manually change every bit 1 to 0. Random perturbations can be seen as the exploration of the possible paths through which an object may evolve over time.}
\label{programs}
\end{figure}

Fig.~\ref{programs} can help illustrate how uniform random perturbations may provide a picture of the set of possible future states and may be relevant in the case of naturally evolving systems. This means that, in both cases in Fig.~\ref{programs}, the asymmetric cost of moving random to simple and simple to random from a purely algorithmic perspective may also be relevant in the case of naturally evolving systems. The degree to which the perturbations are informative depends on whether the dynamic evolution process is proceeding (nearly) uniformly at random or not. 
An alternative way to illustrate the phenomenon of asymmetric reprogrammability is by considering networks as produced by computer programs (Fig.~\ref{curve}) (or as produced by computer programs). The algorithmic complexity $C$ of a complete graph $k$ grows by its number of nodes because the generating mechanism is of the form ``connect all $N$ nodes'', where $N=|V(k)|$. In contrast, the algorithmic complexity of an algorithmically random Erd\"os-R\'enyi (\textnormal{E-R}) graph with edge density $\sim 0.5$ grows by the number of edges $|E(\textnormal{E-R})|$ because to reproduce a random graph from scratch the sender would need to specify every edge connection, as there is no way to compress the description.

As depicted in Fig.~\ref{curve}, algorithmic-random removal of a node $n$ from a complete graph $k$ produces another complete, albeit smaller graph. Thus the generating program for both $k$ and $k'=k\backslash n$ is also the relationship between $k$ and $k'$, which is of a causal nature. In contrast, if an edge $e$ is removed from $k$, the generating program of $k''= k\backslash e$ requires the specification of $e$ and the resulting generating program of size $C(k'')>C(k)$, sending $k$ towards randomness for every edge $e$ algorithmic-randomly removed. 

On the one hand, moving a complete graph toward randomness (see Fig.~\ref{programs} and ~\ref{curve}) requires algorithmically random changes. On the other hand, we see how a random graph can also be easily rewired to be a complete graph by edge addition. However, if the complete graph is required to exactly reproduce a specific random graph and not just its statistical properties, then one would need to have full knowledge of the specific random graph and apply specific changes to the complete graph, making the process slow and requiring additional information. Yet, moving the random graph 
toward a complete graph requires the same effort as before, because the complete graph is unique, given the fixed number of nodes and implies reaching edge density 1 no matter the order of the added edges. Nevertheless, transforming a simple graph, such as the complete graph (see Fig.~\ref{curve},) by algorithmically random one-by-one edge removal has a greater impact on its original algorithmic complexity than performing the same operation on a random graph.

\subsection{Graphs as Computer Programs}

Specifically, if $S$ is a simple graph and $R$ an algorithmically random one, then we have it that $C(S)-C(S\backslash e)\geq C(R)-C(R\backslash e)$, i.e., the rate of change from $S$ to (a non-specific) $R$ is greater than in the other direction, thus imposing a {thermodynamic-like} asymmetry related to the difficulty of reprogramming one object into another according to its initial program-size complexity. This is because a random deletion to $R$ will always have little impact on the underlying $print(R)$ with respect to $|R|$ (of at most logarithmic effect), because $print(R) \sim R$, but a random deletion (or a random transformation in general) to $S$ may lead to a greater disruption of the small (by definition) generating program of $S$. The asymmetric axis where the highest reprogrammability point can be found is exactly the point at which $C(S)-C(S\backslash e)=C(R)-C(R\backslash e)$ for a specific $S$ and $R$.

\begin{figure}[ht!]
\centering
\includegraphics[width=12cm]{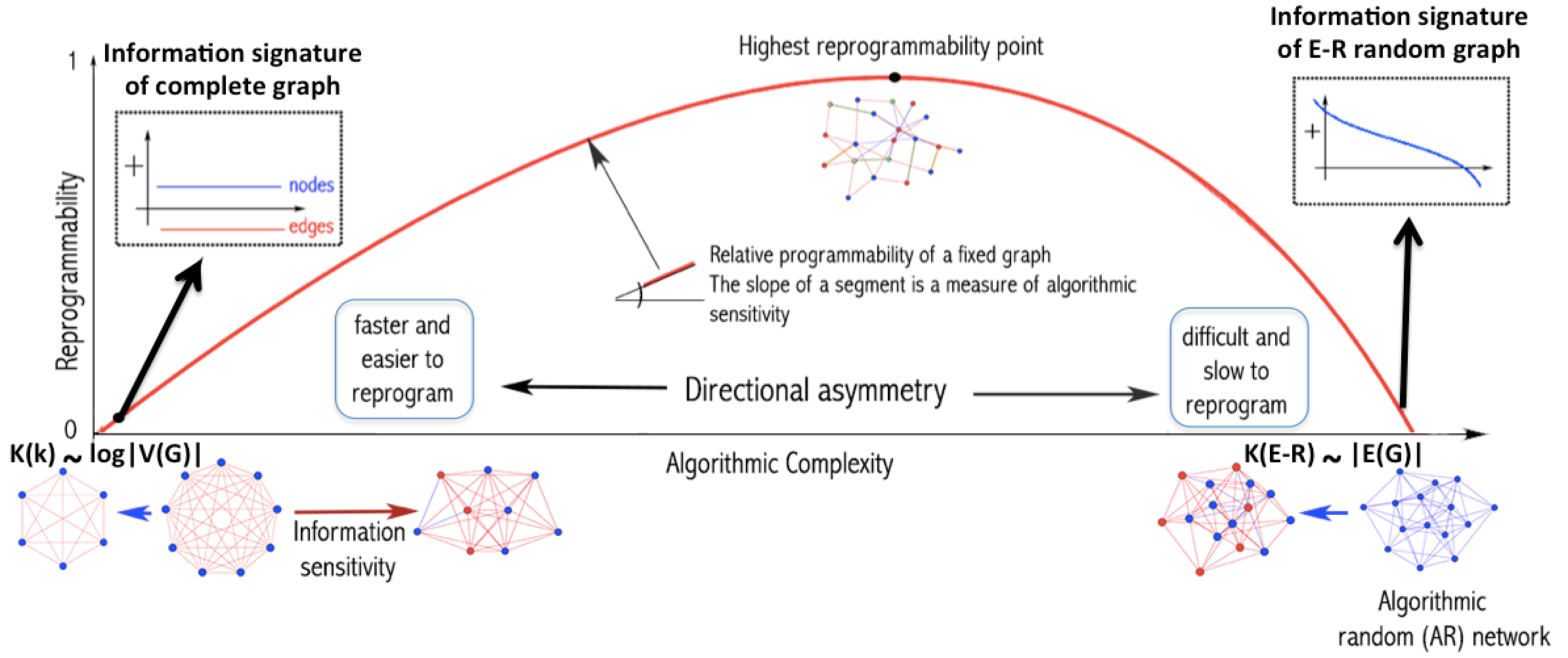}

\caption{Graphs as produced by programs. All real-world networks lie between the extreme cases of being as simple as a complete graph whose algorithmic complexity $C$ is minimal and grows by only $\log|V(k)|$, and a random (also statistically random and thus E-R) graph whose algorithmic complexity is maximal and grows by its number of edges $|E(\textnormal{E-R})|$. If we ask what it takes to change the program producing $k$ to produce $\textnormal{E-R}$ and vice versa, in a random graph, any single algorithmic-random node or edge removal does not entail a major change in the program-size of its generating program, which is similar in size to the random graph itself i.e. $|E(G)|$. The curve shows how, without loss of generality, the reprogramming capability of networks, as produced by computer programs, produces an asymmetry imposed by algorithmic complexity and reminiscent of traditional thermodynamics as based on classical probability. A maximally random network has only positive (blue) elements (Fig. 5) because there exists no perturbation that can increase the randomness of the network either by removing a node or an edge, as it is already random (and thus non-deterministic). Thus changing its (near) minimal program-size length by edge or node removal is slow. However, a simple graph may have elements that incline its program-size length toward randomness. In each extreme case (simple vs random) the distribution of sorted elements capable of shifting in each direction is shown in the form of what we call `signatures', both for algorithmically random edge and node removal. The highest reprogrammability point is the place where a graph has as many elements to steer it in one direction as in the other.}
\label{curve}
\end{figure}

It is clear then how analysing the contribution of each element to the object, as shown in Fig.~\ref{curve}, has the potential to reveal the algorithmic nature of the original object and how difficult it is to reprogram the underlying generative computer program in order to produce a different output/graph.

A thermodynamic-like effect can be found in the (re)programmability capabilities of an object. Moving random networks by edge removal is significantly more difficult than moving simple networks towards randomness. For random graphs, there are only a few elements, if any, that can be used to convert them slowly towards simplicity, as shown in Fig.~\ref{curve}. In contrast, a larger number of elements can nudge a simple network faster towards randomness. This relationship, captured by the reprogrammability rate for simple versus random graphs, induces a thermodynamic-like asymmetry based on algorithmic complexity and reprogrammability. 

Fig.~\ref{curve} illustrates that in a complete graph, the algorithmic-random removal of any single node leads to a less than logarithmic reduction in its algorithmic complexity, while the removal of any single edge leads to an increase in randomness. The former because the result is simply another complete graph of a smaller size, and the latter because an algorithmically random deleted link would need to be described after the description of the complete graph itself. In other words, If $k_n=$complete graph on $n$ nodes, $C(k_n)$ is at most $C(n)+O(1) \leq log_2(n) + O(1)$ versus $C(k_{n-1}) \leq k(n-1) + O(1) \leq log_2(n-1) + O(1)$, so the difference between $log_2(n)$ and $log_2(n-1)$ is very small and not even the difference between,
say, $n$ and $n-log_2(n)$, which is what one would normally call a ``logarithmic additive difference''.

It is worth noting that if we only want to delete a single edge from the complete graph, there are $|V(k_n)| \choose 2$ possibilities . However, there is only one such possibility up to isomorphism. So if we only care about describing graphs up to isomorphism (or unlabelled), the complexity drops only by a constant. This is true for deleting any $n$ edges, as there will then only be $n$ possible isomorphisms. For this and the general case, we have previously investigated the relationship between labelled and unlabelled graph measures of algorithmic complexity~\cite{zenilgraph,kolmo2d,methodszenil}, and while there are some differences, for the most part the results hold, though they warrant some minor variations.

If a graph is evolving deterministically over time, its algorithmic complexity remains (almost) constant up to a logarithmic term as a function of time, because its generating mechanism is still the same as in the complete graph (which is a simplest extreme case), but if its evolution is non-deterministic and possible changes to its elements are assumed to be uniformly distributed, then a full perturbation analysis can simulate their next state, basically applying exhaustively all possible changes one step at a time. In this case, any node/edge perturbation of a simple graph has a very different effect than performing the same interventions on a random graph.

\section{Principle of Maximum Algorithmic Randomness (MAR)}

There is a wide range of applications in science, in particular in statistical mechanics, that serve as ways to study and model the typicality of cases and objects (see Discussion section). Indeed, determining how removed an object is from maximum entropy has been believed to be an indication of its typicality and null model based on its information content~\cite{bianconi}.

Maximum entropy, or simply Maxent, is taken as the state of a system when it is at its most statistically disordered---when it is supposed to encode the least information. 

\begin{figure}[ht]
\centering
\includegraphics[width=9cm]{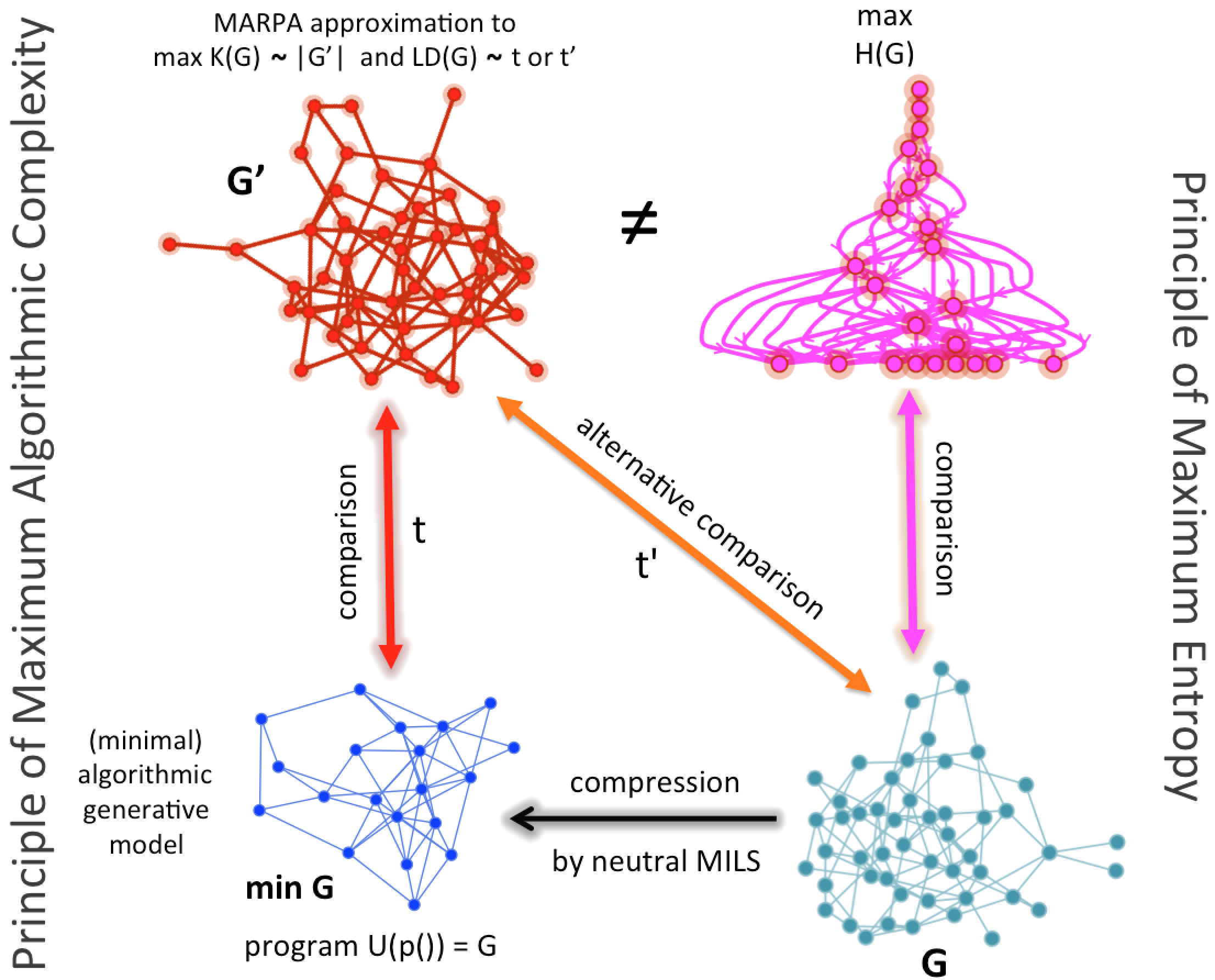}

\caption{Algorithmic Maxent in the generation of graphs (without loss of generality, no other restrictions are imposed). The paths to statistical and algorithmic randomness are different, and they determine different principles for different purposes. Algorithmic Maxent tells apart recursive cases from non-recursive (thus algorithmic random) ones. Classical Maxent quantifies statistical randomness but its algorithmic refinement quantifies both statistical and algorithmic randomness. This opens up the range of possibilities for moving toward and reaching a random graph, by not only considering whether it is random-looking but also whether it is actually algorithmically random. Because the worse case for algorithmic Maxent is not to distinguish recursive cases (due to e.g. semi-computability) its worse performance is to retrieve the same Gibbs measure and produce the same ensemble as classical Maxent would do.}
\label{maxent2}
\end{figure}

We will see here, however, that entropy may collapse cases that are only random-looking but are not truly so. Based on the ideas above we can, nevertheless, suggest a conceptual and numerical refinement of classical Maxent by an algorithmic Maxent. Instead of making comparisons against a maximal entropy graph, one can make comparisons against a generated maximal algorithmic random graph (MAR) candidate (shown in red in Fig.~\ref{maxent2}) by either comparing the original graph (denoted by $G$ in Fig.~\ref{maxent2}) or a compressed version (denoted by $minG$) approximated by, for example, algorithmic graph sparsification~\cite{milspaper}. Such comparisons are marked as $t$ and $t'$ in Fig.~\ref{maxent2} and replace the need for comparison with a maximal entropy graph that may not be algorithmically random. Fig.~\ref{maxent2} shows how such a replacement can be made and what comparisons are now algorithmic ($t$ and $t'$) versus only statistical, which in the case of this illustration involves a graph that has been shown to produce a near maximal degree sequence when it is of lowest algorithmic randomness~\cite{zkpaper} (in magenta, top right).

\subsection{Maximal Algorithmic Randomness Preferential Attachment (MARPA) algorithm}
\label{maxent3}

Once the number of nodes is fixed, a MAR graph is of density 0.5, just like a classical E-R graph. This is because the highest algorithmic complexity is reached when $K(G) \sim |E(G)|$ is maximised somewhere between the 2 extreme cases in which fully disconnected and complete graphs reach minimum complexity $K(G) \sim log_2 |V(G)|$.

This means that, without loss of generalisation to other objects, a Maximal Algorithmic Random graph $G$ is an Erd\"os -R\'enyi (\textnormal{E-R}) graph that is algorithmically random, i.e. whose shortest possible computer description is not (much) shorter than $|E(G)|$, where $|E(G)|$ is the number of edges of $G$; or, $|E(G)| - C(G)  < c$.

MARPA seeks to maximise the information content of a graph $G$ by adding new edges (or nodes) at every step. The process approximates a network of a given size that has the largest possible algorithmic randomness and is also an Erd\"os-R\'enyi (ER) graph. An approximation of a `Maximal' Algorithmic- Random (MAR) graph can be produced as a reference object whose generating program is not smaller than the network (data) itself and can better serve in maximum (algorithmic-) entropy modelling.

MARPA allows constructions of a maximally random graph (or any object) by edge preferential attachment, in such a manner that randomness increases for any given graph. Let $G$ be a network and $C(G\backslash e)$ the information value of $e$ with respect to $G$ such that $C(G) - C(G\backslash e) = n$. Let $P = \{p_1, p_2, \ldots, p_n\}$ be the set of all possible perturbations. P is finite and bounded by $P < 2^{|E(G)|}$ where $E(G)$ is the set of all elements of $G$, e.g. all edges of a network $G$. We can approximate the set of perturbations $e'$ in $P$ such that $C(G) - C(G\backslash e') = n'$ with $n' < n$. As we iterate over all $e$ in $G$ and apply the perturbations that make $n' < n$, for all $e$, we go through all $2^{|E(G)|}$ possible perturbations (one can start with all $|E(G)|$ single perturbations only) maximising the complexity of $G' = max\{G | C(G) - C(G\backslash e) = \{\textnormal{max among all $p$ in $P$ and $e \in G$}\}$ (which may not be unique and can build an algorithmic probability distribution to compare with, one that is complementary to the distribution of maximal entropy graphs).

We denote the set of maximal entropy graphs as $\{MaxS(G)\}$ and the set of maximal algorithmic randomness graphs as $\{MaxK(G)\}$, while approximations to $\{MaxK(G)\}$ will be denoted by $\{MaxC(G)\}$.

\begin{figure}[ht]
\centering
\includegraphics[width=10cm]{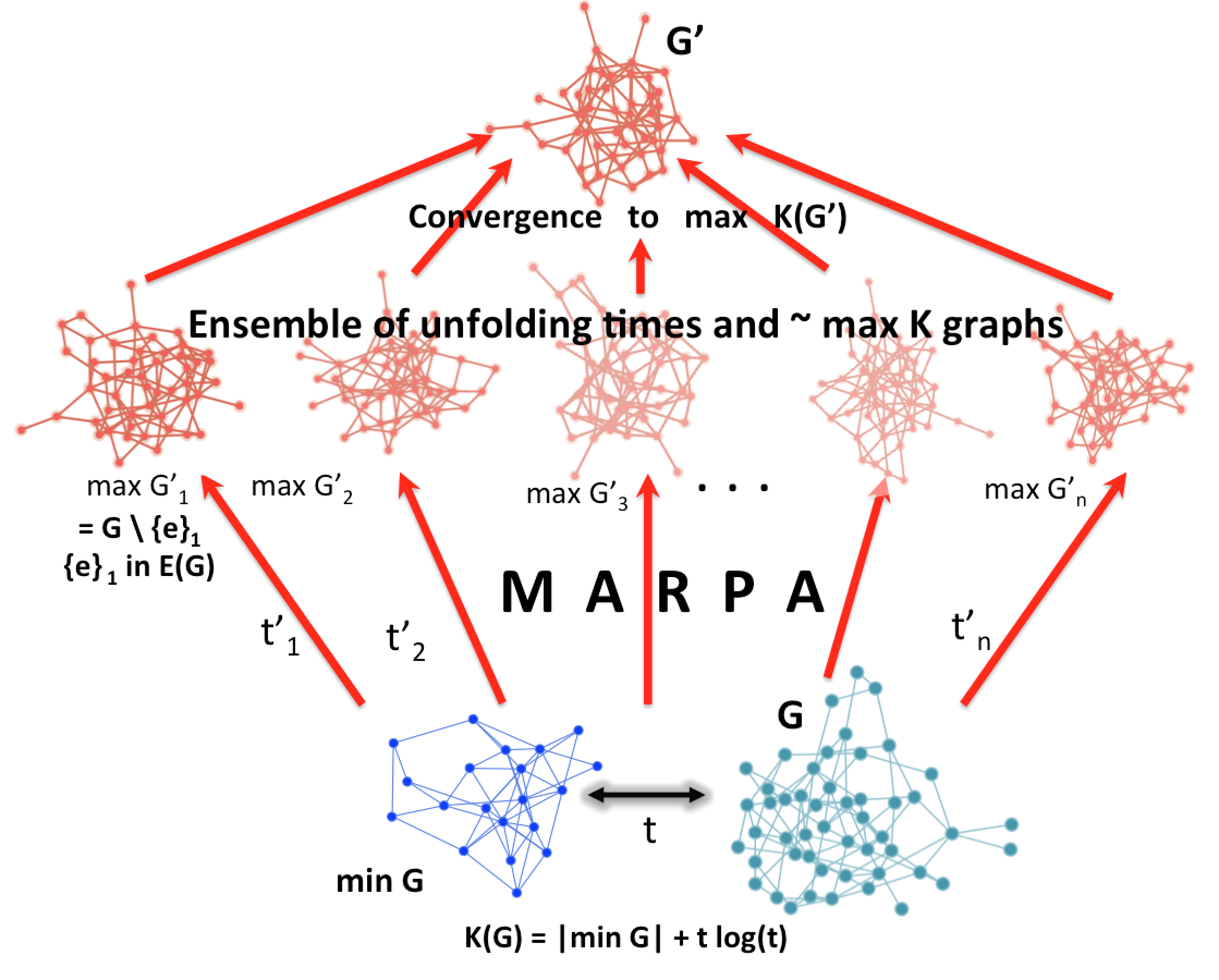}

\caption{Reduction of the Gibbs/Boltzmann distribution. While objects may appear maximally statistically random in all other properties but the ones constrained by a problem of interest for which Maxent is relevant, there may actually be of different nature and some may actually be algorithmically random while others recursively generated hence they should not be placed in the same distribution. The original graph $G$ has a shortest description $min G$ and perturbations to either the original and compressed version can lead to randomised graphs for different purposes. Some of them will have maximum entropy denoted by $max G'_n$ but among them only some will be also of $max K(G)$.}
\label{maxent}
\end{figure}

Approximating methods allow tighter bounds for MAR graphs beyond entropy approaches. This algorithmic version expands and improves over classical Maxent (Fig.~\ref{maxent}) which is based on classical information theory and Shannon entropy. While Fig.~\ref{maxent} may be misleading because there is no guarantee of uniqueness of $G'$ such that $K(G')$ is of maximal algorithmic randomness (neither in theory nor practice) under some restriction under classical Maxent, the number of algorithmically random graphs is strictly smaller than the number of high entropy graphs for the same size and under any other constraint (e.g. same degree sequence or edge density) and thus can be seen as a reduction of the ensemble size and a refinement of classical Maxent. 

\begin{definition}
The randomness deficiency of an object, such as a graph, in a particular distribution (e.g. a set of graphs) quantifies how far an object is from the greatest algorithmic random object with the same properties (e.g. edge density) in the distribution.
\end{definition}

The purpose of MARPA (Fig.~\ref{maxent}) is thus to maximise the randomness deficiency between an object of interest and its greatest algorithmic randomisation. To this end, it is necessary to estimate the algorithmic complexity of such an object by means such as popular lossless compression algorithms (e.g. LZW), which are limited but are an improvement over other measures such as entropy~\cite{emergence}, or by alternative means such as those introduced  in~\cite{d4,d5,kolmo2d,bdmpaper,methodszenil} based on algorithmic probability.

Let such a constructed maximal complexity object $maxC(G)$ be used to quantify the randomness deficiency of an object of interest be denoted by $C(G)$. The procedure consists in finding the sequential set of perturbations $\{P\}$ that upon application to $G$ leads to $G'$ such that $maxC(G) - C(G')$ is minimised, and where $C(G') - C(G)$ is the randomness deficiency estimation of $G$, i.e. how removed $G$ is from its (algorithmic-)randomised version $maxC(G)$ (notice that $C(G)$ is upper bounded by $maxC(G)$ and the difference is always positive).

\subsection{Supremacy of algorithmic Maxent}

To illustrate how algorithmic Maxent is an improvement over classical Maxent we demonstrate that the E-R model admits different constructions some of which are not algorithmic random.

\begin{theorem} (existence) At least one E-R graph exists which is  not algorithmically random.
\label{theoexist}
\end{theorem}
Ackermann~\cite{ackermann} and Rado~\cite{rado} constructed the so-called Rado graph in a similar spirit to the ZK-graph~\cite{zkpaper} using an algorithmic/computable procedure based on what is called the BIT predicate by identifying the vertices of the graph with the natural numbers $0, 1, 2, \ldots$ An edge connects vertices $x$ and $y$ in the graph (where $x < y$) whenever the $x$th bit of the binary representation of $y$ is nonzero. So, for example, the neighbours of vertex zero consist of all odd-numbered vertices, because the numbers whose 0th bit is nonzero are exactly the odd numbers. Vertex 1 has one smaller neighbour, vertex 0, as 1 is odd and vertex 0 is connected to all odd vertices. The larger neighbours of vertex 1 are all vertices with numbers congruent to 2 or 3 modulo 4, because those are exactly the numbers with a nonzero bit at index 1. The Rado graph is almost surely Erd\"os–R\'enyi on countably many vertices~\cite{ackermann,rado}.

\begin{theorem} (non-uniqueness) There is more than one E-R graph that is not algorithmically random.
\label{uniqueness}
\end{theorem}
 The set of pseudo-random generating graphs with fixed seeds of different minimum program-size generates one E-R random graph, and for each different seed it will generate another E-R random graph.Therefore there more than one E-R graph that is not algorithmically random.
it should be noted that an  \textnormal{E-R} graph with density 0.5 may be of maximal entropy, but can be produced by programs of varying length and thus of varying algorithmic randomness, i.e. either recursively generated or not.
Fig.~\ref{numericalplot} shows that the complexity of graphs with exactly the same number of nodes and edges that comply with the properties of an E-R graph (e.g. edge density $\sim$ 0.5) do not comply with the property of maximum algorithmic randomness (MAR), and their values will diverge for typical randomly chosen examples of growing graphs by node count. It is also shown that the numerical approximations, both for simple (complete) and MAR graphs, follow the theoretical expectations. The proof is by induction.

\subsection{Numerical examples}

Consider the absolute maximum algorithmic-random graph, which we will denote by $maxC(G)$. $maxC(G)$ is a graph comprising the same number of nodes but with an edge rearrangement operation such that $C(G) < C(max(G)) \leq 2^k$, where $k = (|E(G)|(|E(G)|-1))/4$ is the maximum number of edges in $G$ divided by 2 where at edge density $0.5$ it reaches maximal algorithmic randomness. 

There are two possible methods to produce estimations of MAR graphs on top-down and one bottom-up. The bottom-up method consists in adding edges starting from the empty graph until reaching the desired target network size (in nodes and/or edges). 

The top-down method consists in deleting edges starting from a given network and find the elements that maximise its algorithmic complexity. We call these methods  edge-deletion and edge-addition algorithmic randomizations (as opposed to just randomization in the classical sense). When constraining edge-addition by maximum number of nodes, both methods converge in maximal algorithmic value and statistical properties but may construct different graphs, both of which are estimation of MAR graphs.

A third method may consist in edge rearrangement/switching but this requires an exponentially greater number of operations at every step. A computationally more efficient way to produce an approximation to a MAR network by edge deletion is to start from a complete graph and keep deleting edges until a target network size is reached. The pseudo-code of the edge-deletion algorithmic randomization is as follows: 

\begin{myenumerate}
    \item 
	Start with an empty or complete graph $G$.
    \item 
	Perform all possible perturbations for a single operation (e.g. edge deletion or addition) on $G$ to produce $G'$.
	\item Keep only the perturbation that
	maximized algorithmic randomness, i.e. $C(G') \geq C(G)$.
	\item Set $G := G'$
	\item Repeat 1 until final target size is reached or $C(G)>C(G')$.
\end{myenumerate}

With the cost function maximizing algorithmic complexity the above pseudo-code is valid for other operations to $G'$ Without loss of generalisation. These operations can be of other types, including edge rearrangement/switching and edge addition just as edge removal (see Discussion). However, different operations require different time complexities.

The edge-addition algorithmic randomization  can start from an empty set, for which one would be generating approximations of MAR graphs (for that operation) for all sizes from bottom up.

The algorithm can also be repeated to produce a set of candidate MAR graphs (not only because many MAR graphs may exist but because $C$ is semi-computable and thus  they are only approximations). The intersection of both $\{MaxC(G)\}$ and $\{MaxS(G)\}$ is an interesting candidate set with which to replace $\{MaxS(G)\}$ alone for purposes of Maxent. But if the algorithm to produce $\{MaxC(G)\}$ approximates $\{MaxK(G)\}$ then $\{MaxS(G)\} \subset \{MaxC(G)\}$ and $\{MaxS(G)\}$ will be upper bounded by $\{MaxC(G)\}$ because it is upper bounded by $\{MaxK(G)\}$.

\begin{figure}[ht!]
\includegraphics[width=4.5cm]{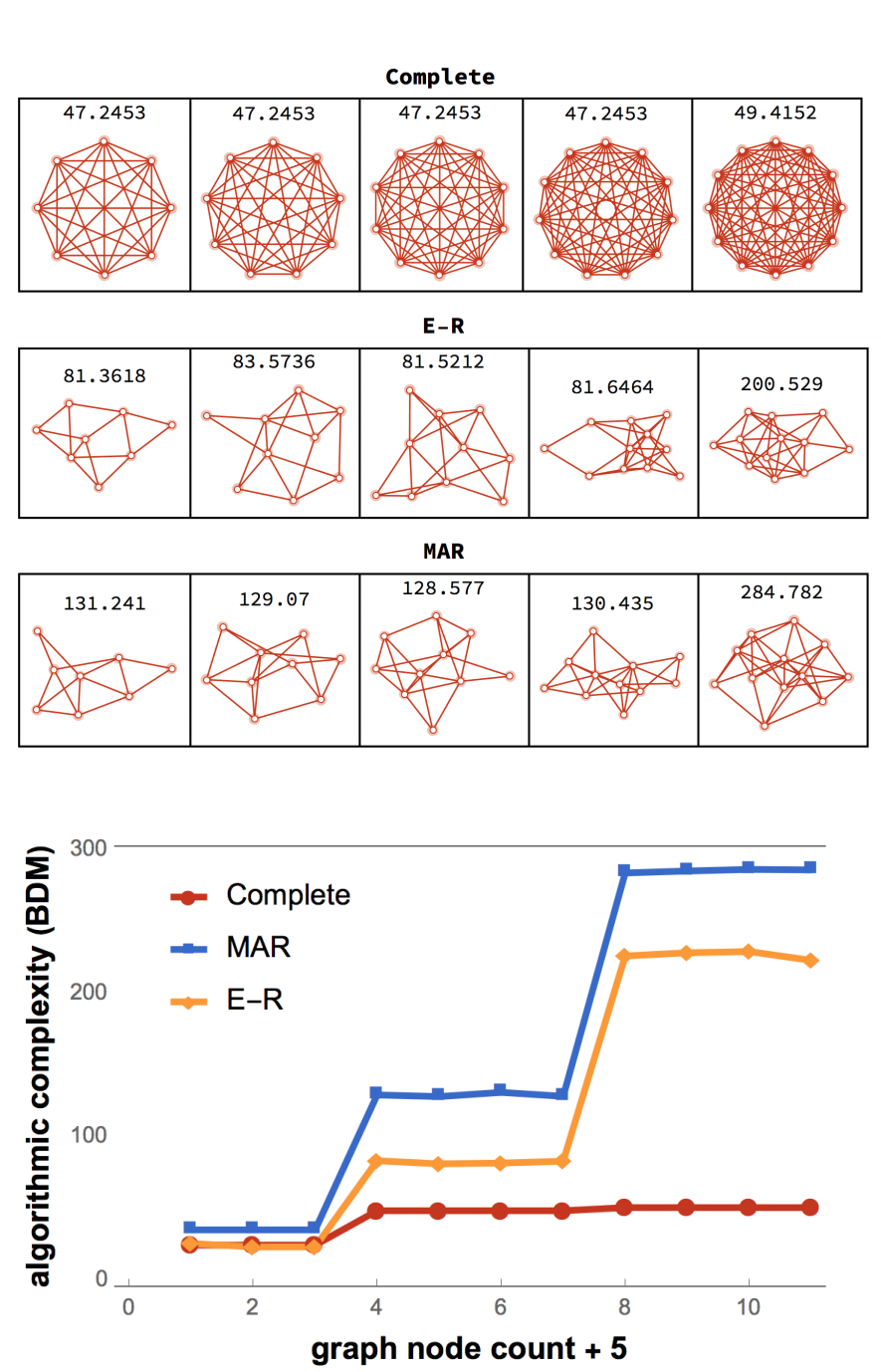}\hspace{.4cm}\includegraphics[width=7.4cm]{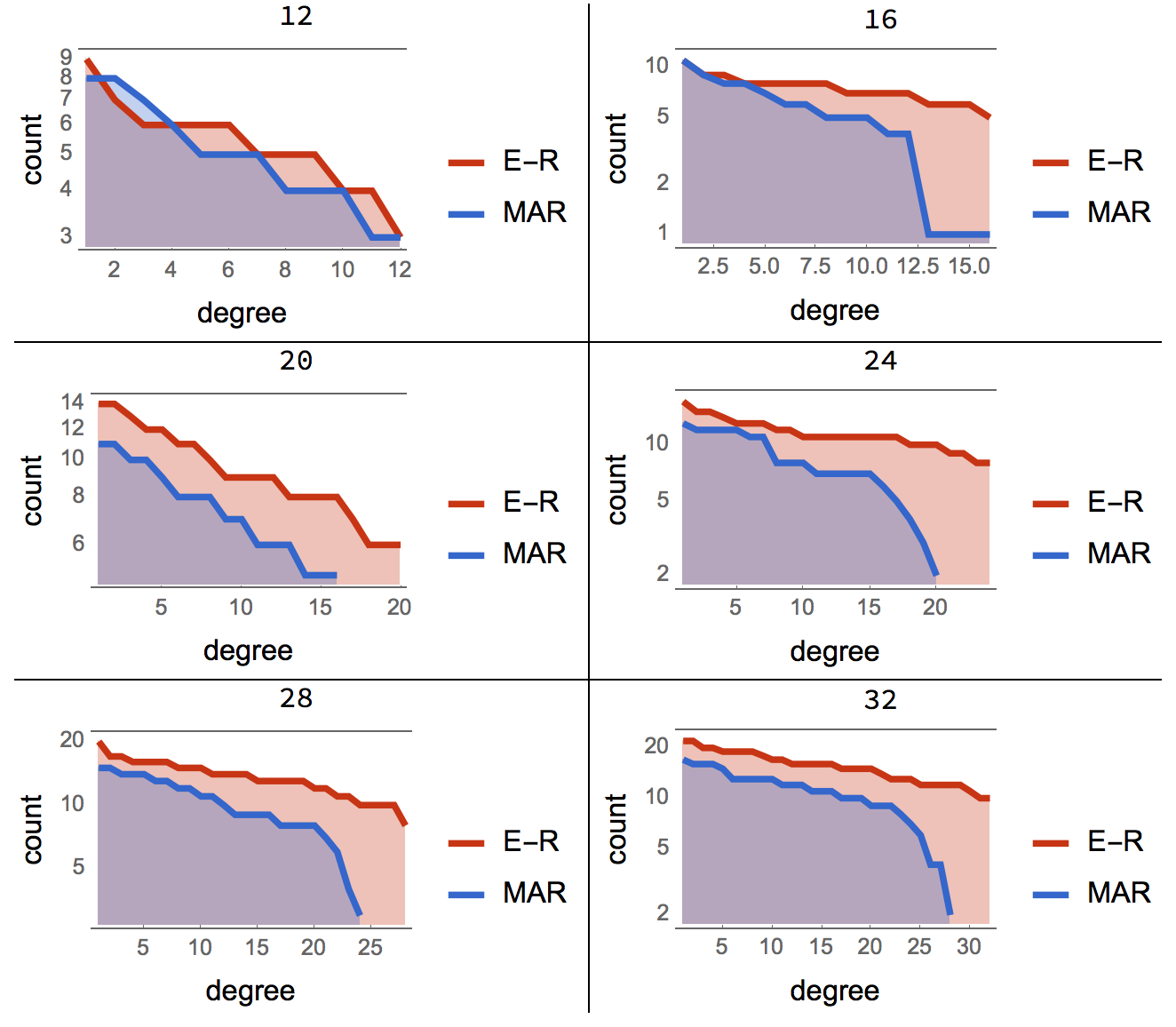}
\label{numericalplot}
\caption{Top left: One can produce a MAR graph starting from an empty graph and adding one edge at a time (see Fig.~\ref{randomization}) or one can start from a complete graph and start deleting edge by edge keeping only those that maximize the algorithmic randomness of the resulting graph. Bottom left: Following this process, MAR graphs top E-R graphs meaning BDM effectively separate low algorithmic complexity (algorithmic randomness) from high entropy (statistical randomness), where entropy would simply be blind collapsing all recursive and non-recursive cases. Right: degree distribution comparison between E-R and MAR graphs.}
\end{figure}

Fig.~\ref{numericalplot} shows how MAR graph degree distributions follow E-R graphs for nodes with high degree but not for lower degree nodes. MAR graphs produce a degree sequence of greater entropy and algorithmic complexity because an algorithmic random graph should also have an algorithmic random degree sequence.

\begin{figure}[ht!]
\centering
\includegraphics[width=8cm]{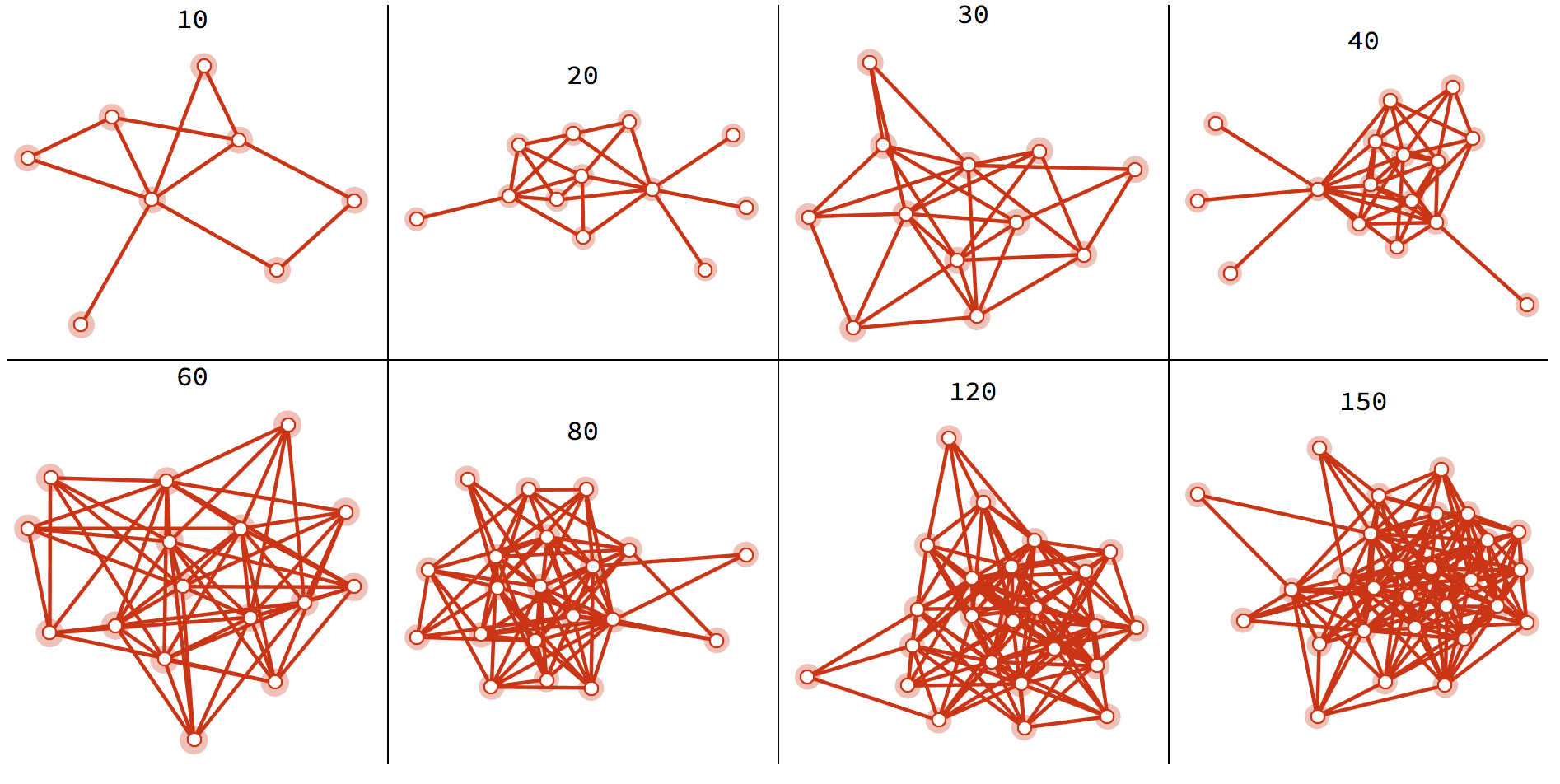}

\bigskip

\includegraphics[width=5.7cm]{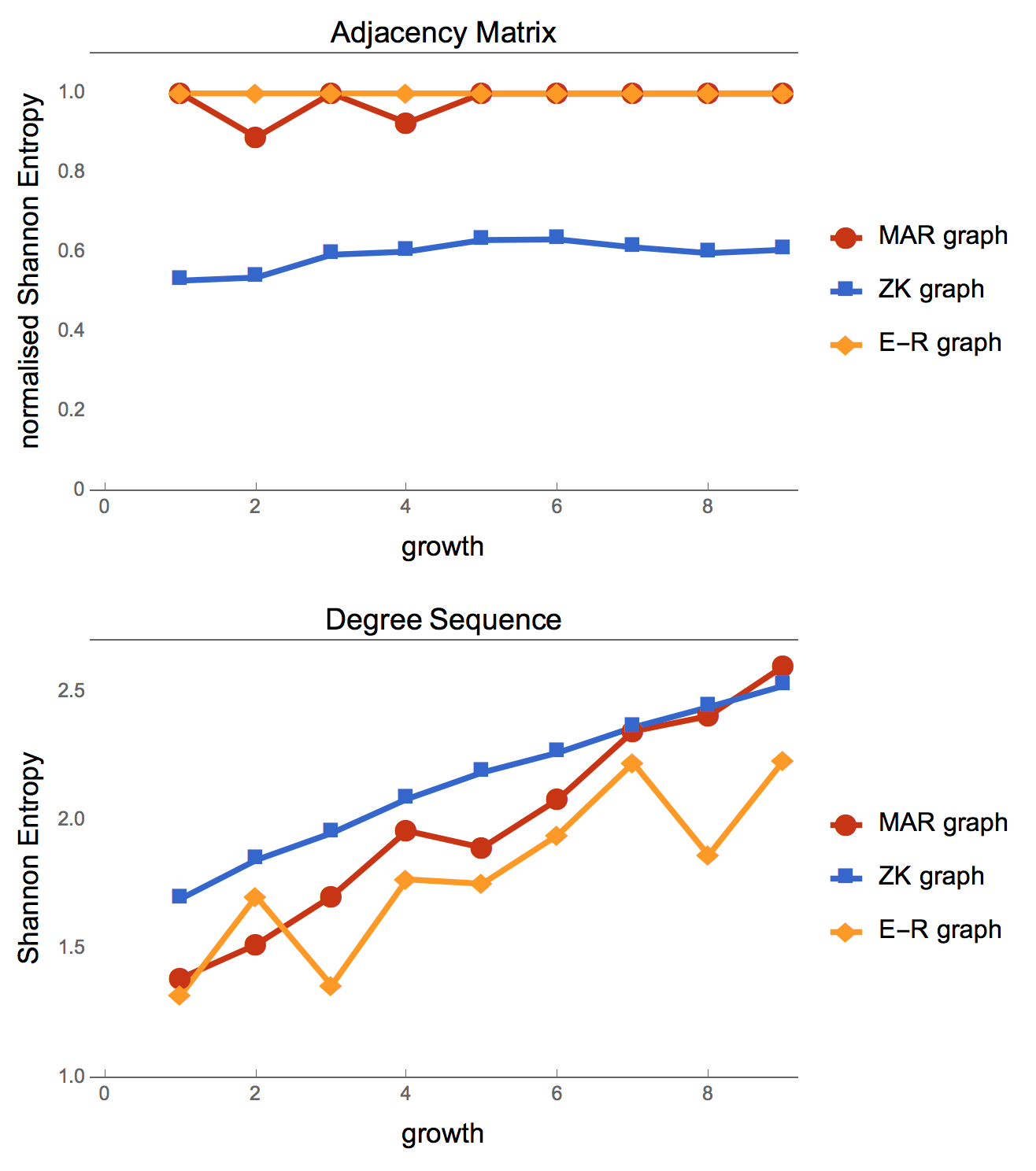}\hspace{.5cm}\includegraphics[width=5.7cm]{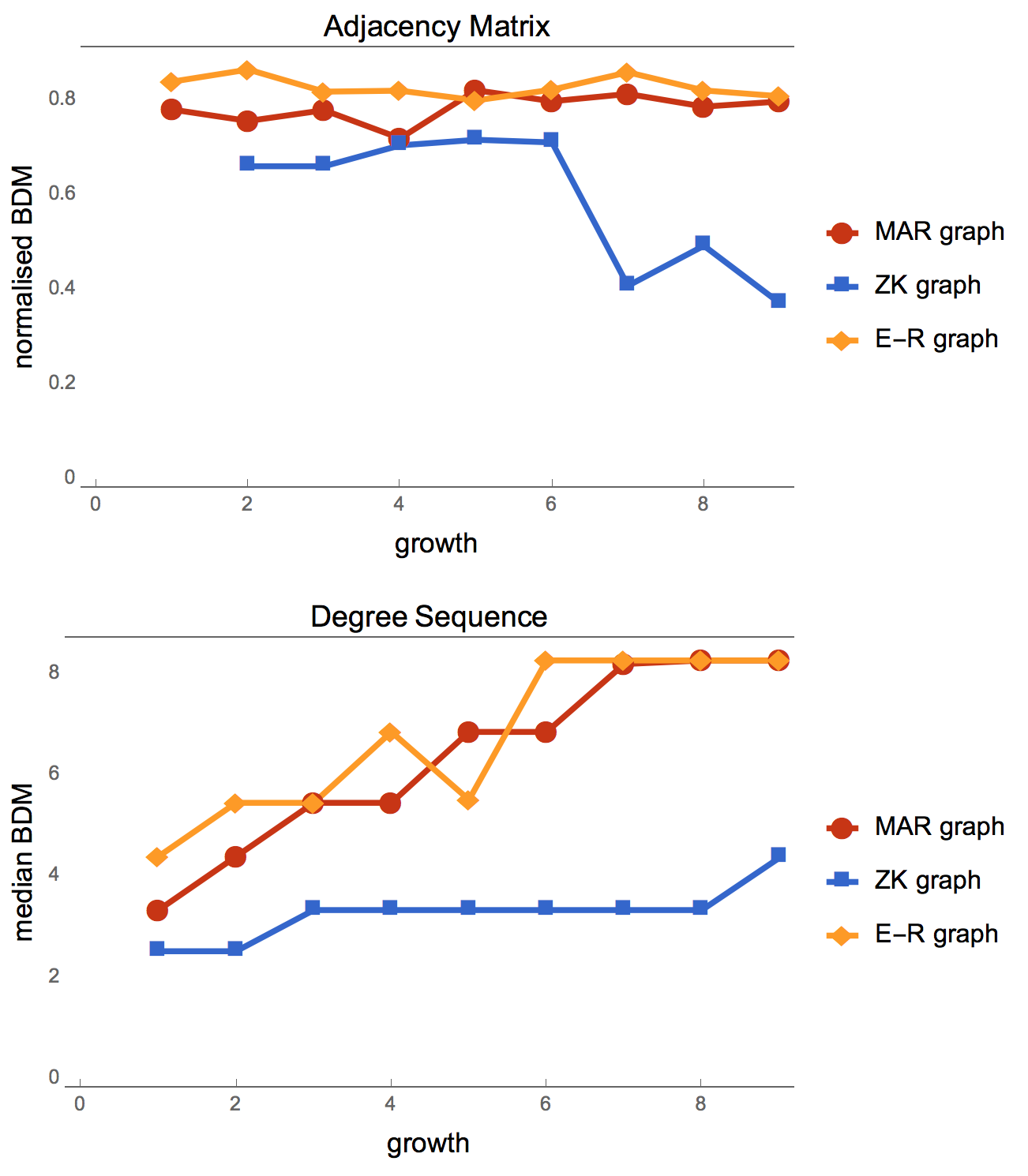}
\caption{Top: A MAR graph constructed by adding one by one every possible edge and keeping only those that maximise the uncompressibility/ algorithmic complexity of the graph according to BDM. Shown are only eight steps from 10 edges to 150. Bottom: MAR graphs versus ensemble of pseudo-randomly generated E-R graphs (with edge density 0.5) versus the ZK graph~\cite{zkpaper}  designed to have minimal adjacency matrix entropy but maximal degree sequence entropy.}
\label{randomization}
\end{figure}

In Fig.~\ref{randomization}, the ensemble of E-R graphs are a randomization of the MAR graphs with the same number of nodes. Despite allowing any possible number of edges, MAR graphs are equally random both according to classical and algorithmic (estimations) on their adjacency matrices but of greater entropy than E-R graphs. The ZK graph came high for degree sequence entropy as expected as it was designed for that purpose, but for approximations of algorithmic complexity by BDM, the ZK graph was considerably less random also as theoretically expected (as it was recursively generated~\cite{zkpaper}. MAR graphs, however, had both maximal entropy and maximal algorithmic randomness.

The edge density of the MAR graphs remains at 0.4 meaning that the greatest increase of algorithmic randomness is achieved by adding new nodes at about 0.4 density rather than increasing the density. When forcing to keep the same number of nodes and reach maximal algorithmic randomness the algorithm does produce MAR graphs with 0.5 edge density.

Attempts to produce MAR graphs with popular lossless compression algorithms (Compress and Bzip2) failed because they were not sensitive enough to small changes required to evaluate the edge to add or delete to generate an approximation of a  MAR graph.

\subsubsection{Time Complexity and Terminating Criterion}

Approximating a MAR graph candidate can be computationally very expensive, with exponential time complexity in the order $O(2^{n^2})$ for the bottom-up approach (adding edges) because at every step, all possible additions have to be tested and evaluated against the original object. The time complexity is, however, reduced to linear time $O(n)$ for the top-down method (starting from a complete graph) when only considering single-perturbation analysis, that is, instead of applying all possible subsets only individual ones are allowed. 

Small MAR graphs can also be produced and are not necessary to recompute once the number of nodes (and possibly number of edges) is fixed. Better classical randomizations are also possible at reasonable time complexity especially because they are only applied to graphs of fixed size and thus asymptotic time constraints are less relevant.

The terminating criterion for the top-down approach to produce MAR graphs consists in stopping the method either at the desired size by edge count or when further deletion leads to a decrease in complexity. In other words, starting from e.g. a complete graph $k$, if $C(k)>C(k'^n)$ where $n$ denoted the $n$ iteration of the method description, then stop at iteration $n-1$ that maximizes algorithmic randomness for all $n$. Otherwise said $C(k)>C(k'^n)$ is concave as a function of $n$ and the terminating time is at the global maximum of $C$.

\section{Discussion}

 We have introduced a principle of maximum entropy (Maxent) based on computability, leading to a stronger principle based on algorithmic randomness, an approximation to a generalisation of the traditional \textit{Maxent} based instead on algorithmic complexity approximations which are independent of the specific method such as the nature of the lossless compression algorithm.
 
 Unlike E-R graphs, MAR graphs cannot be generated by computer programs of much smaller size than the edge count of the networks themselves. The intuition behind the construction of a MAR graph is that the shortest computer program (measured in bits) that can produce the adjacency matrix of a MAR graph, is of about the size of the adjacency matrix and not significantly shorter.  
 

We believe that deconvolving aspects of randomness versus pseudo-randomness in scientific practice can help elucidate many confounding aspects within certain areas such as physics. For example, the so-called \textit{Holographic Principle}, where it is often suggested that `all the information about the universe is contained in its surface', but makes no distinction between statistical and algorithmic information.

Roughly, if the universe has positive entropy, then the set of universes with the same probability distribution is much larger than the number of actual universes in which we live. For example, for a universe statistically random there are many more statistical-equivalent candidates, including those algorithmically generated reproducing the same statistical distribution (for an example of how an object can mimic to have a target probability distribution, appear random but not being so see~\cite{zkpaper}), than candidate universes of low entropy universe. This is because for only for low entropy candidate universes will be also of low algorithmic randomness producing a small set of possible models, but for higher entropy cases, the number of high entropy but low algorithmic complexity universes diverge. In other words, there are about $Q^{n-c}$ ways to describe $Q$ particles contained in a universe of length $n$ with less than $c$ bits than $Q/S_Q!$ many ways to place $Q$ particles in the same space, preserving the same statistical distribution (even under all coarse-graining partitions $S_Q$). $S_Q$ is the subset of particles $Q$ whose permutation in the same subset preserve entropy $S$. This is because we know that the number of objects in a set with highest algorithmic complexity is a proper subset of the number of objects with highest entropy in the same set. Theo.~\ref{theoexist} and Cor.~\ref{uniqueness} are here relevant as they show this for a specific type of object, showing that there are E-R graphs that have the same properties but can be differently generated by recursive means. In other words, only a fraction of those would actually describe the exact configuration of our universe or some algorithmic equivalent one. The difference is huge considering that the particles that make a human being can be arranged in so many configurations other than a human being (let alone a particular one) preserving the same statistical properties of their original arrangement in a human being. Yet, it is highly algorithmically probable to define a configuration closer to a human being (because it is of low algorithmic complexity, and according to the coding theorem is algorithmically highly probable~\cite{cover}), or even a particular one, when describing the process that led to that human being rather than simply describing its statistical properties.
 
The point is thus that information claims about statistical properties of objects is way less precise than what they apparently convey which is not enough to reconstruct the same object, such as our universe, or its particular generating code. A clear distinction should thus be made between statistical versus algorithmic information and statistical versus algorithmic randomness.

This algorithmic approach provides such a platform, independent of whether we can numerically or methodology achieve such distinction, we claim we can approximate. Our framework offers an alternative to translate and reinterpret generic examples from classical thermodynamics in the language of computer programs. For example, Maxwell's demon is captured or reformulated by a recursively enumerating process keeping record of algorithmic-random deletions, allowing reversing these deletions into non-algorithmic random ones and incurring in an unavoidable cost. The undeletion process then becomes computable by exchanging information loss with memory (e.g. a reverse lookup table), assigning an index to each process from a preconceived, well-defined enumerating scheme, making the deletion process reversible without algorithmic information loss. The incurred cost is in the recording of the data and the preconception of the computable enumeration from which the knowledge is extracted and recorded at every stage back and forth, as their answers can then be seen as coming from an Oracle machine (the algorithmic Maxwell demon) providing the answer to the question of which element is deleted at any given time to make a network more algorithmically random at no cost (`hotter'). The enumeration would then allow us to trace back and forth the location of every element independent of its cause and effect, breaking the asymmetry and apparently violating the asymmetry found in the otherwise natural thermodynamics of computer programs, as described here. 

In the present work we have only considered deletion (of edges or nodes) from a graph but this is without loss of generalisation. Indeed, all the results hold when considering other recursive vs non-algorithmic transformations, such as swapping nodes or edges or swapping bits, rows and columns on its adjacency matrix. The algorithmic asymmetry in reprogrammability will come from whether such a specific transformation is computable (recursive) or algorithmically random. Indeed, the algorithmic information content properties of $G$ are essentially identical to those
of $t(G)$ if $G=t'(t(G))$, i.e. if the transformation $t$ is enumerable and reversible but not if $t$ is non-computable or algorithmically random. Edge deletion is, however, the most interesting approach in practice for cases such as sparse graphs, graph dimension reduction and network reconstruction.

Evidently, our results, methods, and Maxent refining algorithm will rely entirely on the numerical approximations to $K(G)$, and thus on how good the estimation $C(G)$ is. We have introduced in ~\cite{d4,d5} novel methods that have yielded better estimations than those methods based on popular lossless compression algorithms~\cite{bdmpaper,emergence} and on this basis we have discovered what we believe to be interesting applications~\cite{nature,ploscompbio,finite}. For example, the use of popular lossless compression algorithms will make $\{maxS\} \cap \{maxC\} = \{maxS\} \cup \{maxC\}$ for most cases, where $C$ is, for example, LZW, because LZW is closer to entropy $S$ than to algorithmic complexity $K$~\cite{emergence,smalldata}. A review of alternative measures both for entropy and algorithmic complexity is available in ~\cite{review}.

\section{Conclusion}

From an asymmetric relationship in a measure of reprogrammability, we conclude that the probabilistic basis of the second law of thermodynamics has strong similarities and connections to the difficulty of formally repurposing certain objects to induce them to behave in different computational ways. 

We have thus established interesting parallels between computer programming as based on the dynamics of computer programs being repurposed and related to {thermodynamic-like} phenomena in terms of algorithmic probability, and algorithmic complexity, from which reprogrammability indexes can be defined. The lack of symmetry in the operation of rewiring networks with regards to algorithmic randomness implies a direction or an asymmetry which indicates a natural direction for digital thermodynamics. 
After showing that the set of maximum entropy graphs is not the same as the set of maximum algorithmically complex graphs, we introduced a principle akin to maximum entropy but based on algorithmic complexity. The principle of maximum algorithmic randomness (MAR) can therefore be viewed as a refinement of Maxent as it is a proper subset of algorithmically random graphs, having also the highest entropy.

We believe that only by rigorously selecting models from data can we hope to make progress in understanding first principles and that current statistical approaches leave all the final interpretation to the observer and help little or only indirectly in truly exhibiting the properties that such approaches are supposed to grasp or reveal.

\section*{Acknowledgements}

H.Z. wishes to acknowledge the support of Swedish Research Council (Vetenskapsr\r{a}det) grant No. 2015-05299. J.T. acknowledges support from King Abdullah University of Science and Technology.

\newpage

\end{document}